\documentclass[]{aa}  
\usepackage{amsmath}
\usepackage{amssymb}
\usepackage{natbib}
\usepackage{graphicx}
\usepackage{txfonts}
\usepackage[english]{babel}
\usepackage{hyperref}
\renewcommand{\matrix}[1]{\mathbf{#1}}
\renewcommand{\vec}[1]{\mathbf{#1}}

\begin{document}
\title{The infrared emission spectra of compositionally inhomogeneous aggregates composed of irregularly shaped constituents}

\titlerunning{Emission spectra of inhomogeneous aggregates composed of irregularly shaped constituents}

\author{
	M. Min\inst{1}
		\and
	J.~W. Hovenier\inst{1}
		\and
	L.~B.~F.~M. Waters\inst{1,2}
		\and
	A. de Koter\inst{1}
}

\offprints{M. Min, \email{mmin@science.uva.nl}}

\institute{
Astronomical institute Anton Pannekoek, University of Amsterdam,
Kruislaan 403, 1098 SJ  Amsterdam, The Netherlands
	\and
Instituut voor Sterrenkunde, Katholieke Universiteit Leuven, Celestijnenlaan 200B, 3001 Heverlee, Belgium
}

   \date{Received September 15, 1996; accepted March 16, 1997}

 
  \abstract
   {}
   {In order to deduce properties of dust in astrophysical environments where dust growth through aggregation is important, knowledge of the way aggregated particles interact with radiation, and what information is encoded in the thermal radiation they emit, is needed. The emission characteristics are determined by the size and structure of the aggregate and the composition and shape of the constituents. We thus aim at performing computations of compositionally inhomogeneous aggregates composed of irregularly shaped constituents. In addition we aim at developing an empirical recipe to compute the optical properties of such aggregates in a fast and accurate manner.}
   {We performed {CDA} computations for aggregates of irregularly shaped particles with various compositions. The constituents of the aggregate are assumed to be in the Rayleigh regime (i.e. they are much smaller than the wavelength of radiation inside and outside the constituents), and in addition we assume that the dominant interaction of the aggregate constituents is through dipole-dipole interactions. We computed the spectral structure of the emission efficiency in the $10\,\mu$m region for aggregates with 30\% amorphous carbon and 70\% silicates {by volume} with various fractions of crystalline and amorphous components.}
    {We find that the spectral appearance of the various components of the aggregate are very different and depend on their abundances. Most notably, materials that have a very low abundance appear spectroscopically as if they were in very small grains, while more abundant materials appear, spectroscopically to reside in larger grains. We construct a fast empirical approximate method, {based on the idea of an effective medium approximation,} to construct the spectra for these aggregates which almost perfectly reproduces the more exact computations. This new method is fast enough to be easily implemented in fitting procedures trying to deduce the dust characteristics from astronomical observations.}
   {}


   \maketitle
%

\section{Introduction}

Dust grains in many astronomical environments are assumed to be aggregates of small particles of different sizes and compositions. This is especially true for cometary dust and dust in disks around young stars. The dust grains in these environments are formed by coagulation of smaller particles, leading to aggregated structures. In order to interpret observations of light scattered, emitted or absorbed by these composite particles, a method is required that takes the detailed properties of the particles into account. 

Spectral resonances in the refractive index of the particle material translate into resonances in the absorption (and emission) spectra. Besides composition, the spectral location of such resonances is also influenced by the shape and size of the particles. Spectra computed for homogeneous spherical particles may often display resonances with strengths and spectral locations not found in the spectra of irregularly shaped particles \citep[see e.g.][]{2001A&A...378..228F, 2003A&A...404...35M} or astronomical observations \citep[see e.g.][]{2002A&A...390..533H, 2003A&A...401..577B, 2003ApJ...595..522M}. Much progress has been made on computational methods to calculate the optical properties of aggregates of spheres \citep[see e.g.][]{1996JOSAA..13.2266M}. However, aggregates of homogeneous spherical particles {are still likely to display resonances typical for homogeneous spheres, and thus behave poorly when comparing to observations of natural particles. Test computations we performed confirm this.} Therefore, a method is required for computing the optical properties of aggregates of irregularly shaped constituents. {So far, an efficient and widely employable computational technique is still lacking.}

The aggregates expected to be present in astronomical environments like cometary comae or protoplanetary disks are generally assumed to be composites of various materials. {Although the inhomogeneous nature of the particles can be taken into account in several ways} \citep[see e.g.][]{2003ApJ...595..522M, 2006A&A...456..535S, 2007A&A...463.1189K}, such inhomogeneous structures are usually modeled by assuming separated homogeneous particles of various compositions \citep[see e.g.][]{2006ApJ...646.1024H, 2005Icar..179..158M, 2005A&A...437..189V, 2003A&A...401..577B, 1999P&SS...47..773B}. Although widely used, the validity of this approach is not obvious. In addition it is difficult to interpret the sizes and shapes derived for the dust grains in this way. Therefore, in this paper we study the effects of aggregation of various materials in single aggregates on the spectral appearance of these particles. In particular we will look at how the pronounced spectral structure of crystalline silicates is displayed when they are embedded in a larger aggregate.

In section~\ref{sec:computational approach} we outline the computational approach we employ to compute the absorption spectra of inhomogeneous aggregates of irregularly shaped particles. Results of calculations are presented in section~\ref{sec:results}. An efficient empirical method to compute the spectra for these aggregates using an effective medium approach is presented in section~\ref{sec:effective medium}. Finally, in section~\ref{sec:conclusions} we present the conclusions of our study.

\section{Computational approach}
\label{sec:computational approach}

Particles very small compared to the wavelength both inside and outside the particle, i.e. in the Rayleigh domain, respond to electromagnetic radiation as a single dipole with dipole moment
\begin{equation}
\vec{p}=\alpha\vec{E}_\mathrm{inc},
\end{equation}
where $\alpha$ is the polarizability of the particle and $\vec{E}_\mathrm{inc}$ is the incoming electric field.
The radiation scattered by such a small particle is a dipole field, and the mass absorption coefficient, i.e. the absorption cross section per unit mass, is given by
\begin{equation}
\kappa=\frac{k}{\rho V}~\mathrm{Im}(\alpha).
\end{equation}
Here $V$ is the material volume of the particle, $\rho$ is the density and $k=2\pi/\lambda$ where $\lambda$ is the wavelength of incident radiation.

When we do not have a single small particle, but an aggregate of such particles, the interaction of all these particles with each other has to be taken into account. In other words, the incident field each particle experiences is the incoming field and the field emitted by all other particles in the aggregate. In principle all multipole interactions between the constituents have to be taken into account. This is what is done in methods for computing the optical properties of aggregates of spheres \citep[see e.g.][]{1996JOSAA..13.2266M}. Although this is in principle not impossible, for irregularly shaped constituents the computational demand of such an approach is very high, and at the moment not feasible. If the constituents themselves are very small while most of the interactions take place over relatively large distances, like in a large aggregate of many very small constituents, the dominant interaction is dipole-dipole interaction. The optical properties of such an aggregate of coupled dipoles can be computed using the Coupled Dipole Approximation (CDA). In this approximation each aggregate constituent is considered to interact as a single dipole with a given polarizability. The difference with the well-known Discrete Dipole Approximation \citep[DDA, see e.g.][]{1973ApJ...186..705P, 1988ApJ...333..848D} is that in DDA the volume of the constituents themselves have to be discretized using multiple dipoles. This has to be done in order to account for the shape of the aggregate constituents. In the CDA we account for the shape of the constituents by using for each constituent a polarizability representative of an irregularly shaped grain.
%
The mathematical formulation of both the CDA and the DDA is exactly the same; it describes the interaction of dipoles.
However, the numerical implementation might be slightly different. In the DDA one can choose the dipoles to be located on a rectangular grid. This allows for the use of Fast Fourier Transform (FFT) methods to increase the numerical performance significantly \citep{Goodman1991, Hoekstra}. Since the constituents of an aggregate are usually not located on a rectangular grid, this is in general not possible when using the CDA. In addition, the speedup in DDA using the FFT method requires one to fill a rectangular box circumscribing the aggregate with dipoles. Since we wish to consider very fluffy grains with a low volume filling factor, this is very inefficient and would require many additional dipoles. 
A method providing a similar speedup as the FFT method is being developed for the CDA computations using the Fast Multipole Method \citep{Koc2001, Amini2003}.

As we mentioned above the approach of only considering the dipole-dipole interactions is valid if the constituents are small and the long range interactions between them dominate. This is the situation we are currently interested in. Future computations, using a method taking into account the multipole interactions, like the superposition T-matrix method or large scale DDA simulations using multiple dipoles per constituent, can be employed to check the accuracy of the approximation we employ here. {We wish to consider aggregates consisting of up to $\sim 4000$ constituents. In the DDA formulation each dipole has to be discretized by, at least, 100 dipoles. Furthermore, the filling factor of our aggregates is very low, $\sim 5\%$. This taken together implies that for DDA we need $\sim 10^7$ dipoles to compute the aggregates. Although these computations are in principle possible on large supercomputers when using an efficient parallel DDA program, a single wavelength point will still take up to several days computer time. It is therefore not feasible at this moment to check the accuracy of the CDA method using DDA computations.}

When we consider for each particle of the aggregate only dipole interactions we have \citep{1988ApJ...333..848D}
\begin{equation}
\label{eq:DDA}
\vec{P}_j=\alpha_j\left(\vec{E}_{\mathrm{inc},j}-\sum_{k\neq j}^N\matrix{A}_{jk}\vec{P}_k\right),
\end{equation}
where $\vec{P}_j$ is the local dipole moment at the position of particle $j$, $\alpha_j$ is the polarizability of particle $j$, $\vec{E}_{\mathrm{inc},j}$ is the incoming field at the location of particle $j$, and $N$ is the total number of constituents in the aggregate. The matrix $\matrix{A}_{jk}$ determines the electric field at the position of particle $j$ due to the dipole field emitted by particle $k$ \citep[for details see e.g.][]{1988ApJ...333..848D}. When Eq.~(\ref{eq:DDA}) is solved for the $\vec{P}_j$ the mass absorption coefficient can be computed by
\begin{equation}
\label{eq:total kappa}
\kappa=\frac{4\pi k}{\rho V|E_\mathrm{inc}|^2}\sum_{j=1}^N\left\{\mathrm{Im}\left[\vec{P}_j\cdot(\alpha_j^{-1})^*\vec{P}_j^*\right] -\frac{2}{3}k^3\vec{P}_j^*\cdot\vec{P_j}\right\},
\end{equation}
where $V$ is the total material volume of the aggregate, $\rho$ is the average material density, and the asterisks denote the complex conjugates.

The dipole polarizability of a homogeneous spherical particle is given by
\begin{equation}
\alpha=3V\frac{m^2-1}{m^2+2},
\end{equation}
where $m$ is the complex refractive index of the particle material. The polarizability of irregularly shaped particles depends on the particle shape. However, a reasonable approximation for the average polarizability of an ensemble of highly irregular particles is given by the Continuous Distribution of Ellipsoids \citep[CDE; see][]{BohrenHuffman, 2006JQSRT..97..161M}. The CDE polarizability is given by
\begin{equation}
\label{eq:CDE}
\alpha=2V\left[\left(\frac{m^2}{m^2-1}\ln m^2\right)-1\right].
\end{equation}
More realistic polarizabilities can be computed using the method of \citet{2006JQSRT..97..161M}. However, here we wish to study only the effects of using irregularly shaped particles as monomers so we employ the frequently used CDE polarizability for the aggregate constituents.

In the following subsections we will first define two quantities we use for the analysis of the emission spectra of the aggregates. Then we will discuss how the aggregates are constructed and what composition we will assume.

\subsection{Effective mass absorption coefficients}

The mass absorption coefficient as given by Eq.~(\ref{eq:total kappa}) is basically the sum over the absorption contributions of each component of the aggregate. We can thus easily compute the effective mass absorption coefficient of a part of the aggregate by summing over only that part of the components. In this way we can create an \emph{effective mass absorption coefficient} for each material component. The effective mass absorption coefficient of material component $i$ of the aggregate is defined as
\begin{equation}
\label{eq:effective kappa}
\tilde{\kappa}_i=\frac{4\pi k}{f_i\rho_i V|E_\mathrm{inc}|^2}\sum_{j}\left\{\mathrm{Im}\left[\vec{P}_j\cdot(\alpha_j^{-1})^*\vec{P}_j^*\right] -\frac{2}{3}k^3\vec{P}_j^*\cdot\vec{P_j}\right\},
\end{equation}
where $f_i$ is the volume fraction of the total aggregate composed of material component $i$, $\rho_i$ is the material density of this component, and the summation over $j$ now only involves the dipoles composed of material $i$. When there are $N_\mathrm{mat}$ different materials that make up the aggregate, the total mass absorption coefficient is simply given by
\begin{equation}
\kappa=\sum_{i=1}^{N_\mathrm{mat}}\frac{f_i\rho_i}{\rho}~\tilde{\kappa}_i\,\,.
\end{equation}

\begin{figure}[!t]
\centerline{\resizebox{7cm}{!}{\includegraphics{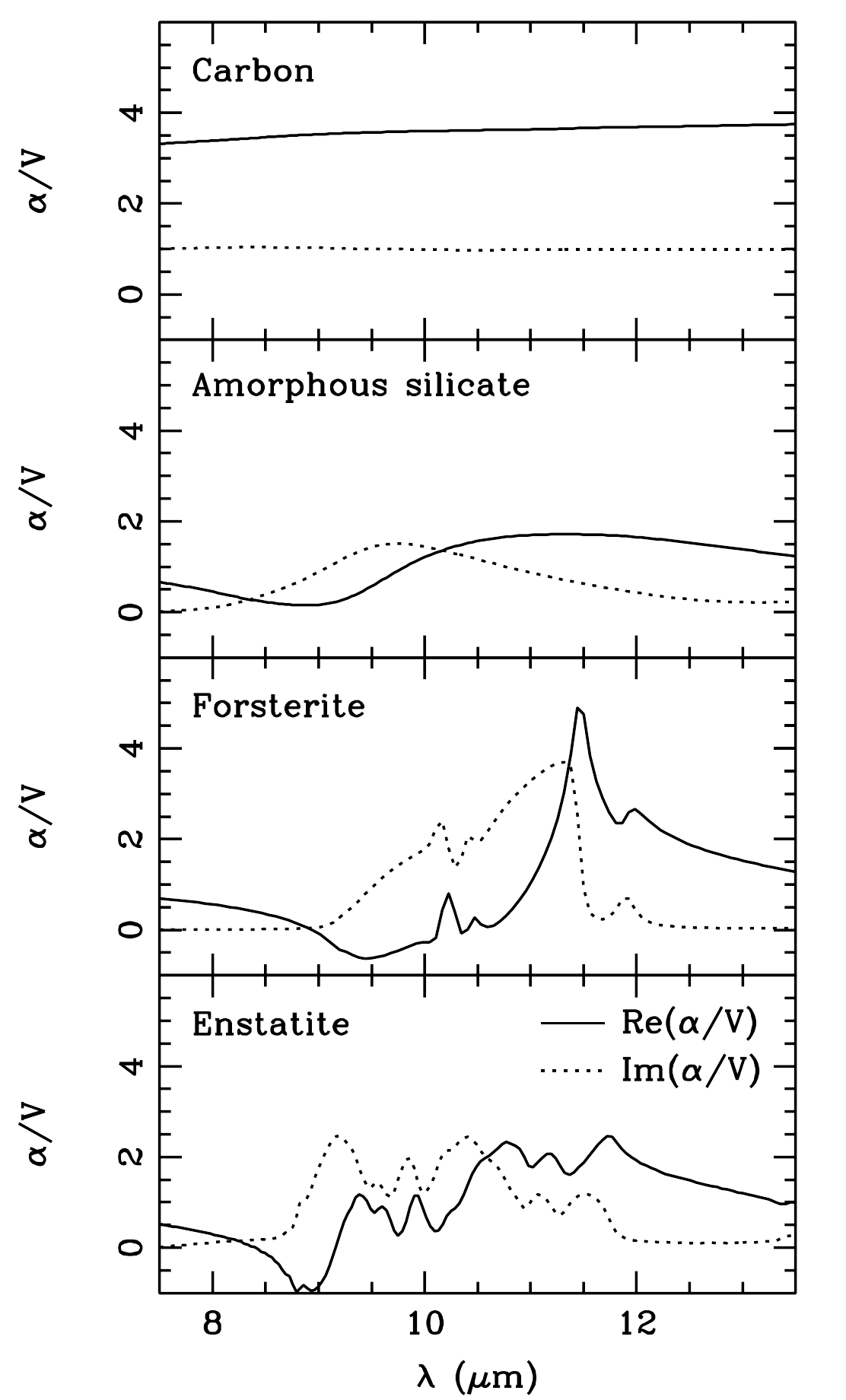}}}
\caption{The real (solid) and imaginary (dotted) part of the polarizability per unit volume for the various components included in the aggregates in the $10\,\mu$m region. For the forsterite and enstatite we display the polarizability per unit volume averaged over the three crystallographic axes.}
\label{fig:alphas}
\end{figure}



\begin{table*}[!t]
\begin{center}
\begin{tabular}{lcccc}
\hline
Name                & Composition   & Lattice Structure     & Volume fraction & Ref.\\
\hline
Amorphous silicate mixture\\
\hline
Fe rich Olivine   & MgFeSiO$_4$       & Amorphous         & 8.6\%	& [1]\\
Mg rich Olivine     & Mg$_2$SiO$_4$     & Amorphous         & 47.7\%	& [2]\\
Mg rich Pyroxene    & MgSiO$_3$         & Amorphous         & 37.0\%	&[1]\\
Na/Al Pyroxene      & NaAlSi$_2$O$_6$   & Amorphous         & 6.7\%	&[3]\\
\hline
Other components\\
\hline
Forsterite          & Mg$_2$SiO$_4$     & Crystalline       && [4]\\
Enstatite           & MgSiO$_3$         & Crystalline       & &[5]\\
Carbon	& C	& Amorphous	& &[6]\\
\hline
\end{tabular}
\end{center}
\caption{The materials used in our aggregates. In the case of the amorphous silicates we use the names 'olivine' or 'pyroxene' to indicate the average silicate stoichiometry. The volume fractions of the amorphous silicates used to construct the average amorphous silicate polarizability are also listed \citep[taken from][]{2007A&A...462..667M}. {Note that these volume fractions represent the relative contributions of these species to the total \emph{silicate} component in the aggregate. Also} note that these fractions are given as volume fractions rather than mass fractions as in \citet{2007A&A...462..667M}. The references in the last column 
refer to:
[1] \citet{1995A&A...300..503D},
[2] \citet{2000A&A...364..282F},
[3] \citet{1998A&A...333..188M},
[4] \citet{Servoin},
[5] \citet{1998A&A...339..904J}, and
[6] \citet{1993A&A...279..577P}.}
\label{tab:Materials}
\end{table*}

\subsection{Construction of the aggregates}

Aggregates in astronomical environments form by coagulation of small grains. The structures that form in this way can have various shapes and degrees of compactness depending on the environmental conditions. On the one hand, fluffy aggregates might form under conditions where aggregates of approximately equal sizes coagulate to form larger structures \citep{kempf99}. On the other hand, very compact structures might form when single monomers are added to a larger aggregate \citep{Ball84}. As shown by \citet{2006A&A...445.1005M} the compactness of the aggregate has a large impact on its spectral appearance. For very fluffy aggregates the spectral structure typical for their constituents is visible, whereas for very compact aggregates the spectral structure of the total aggregate dominates.

We construct the aggregates by adding single particles to a growing aggregate. We start out with a single particle and shoot a second particle with a random impact parameter at the particle till it hits the particle and sticks. The resulting cluster is randomly rotated and a new particle is shot at the aggregate in the same way. In this way we produce moderately fluffy aggregates. In order to construct aggregates of irregularly shaped particles we shoot Gaussian Random Field (GRF) particles at each other \citep{2003JQSRT..78..319G,  2005Icar..173...16S}, thus constructing an aggregate of these very irregularly shaped particles. We then locate a dipole with CDE polarizability at the center of mass of each GRF particle. {The volume equivalent radius of the aggregate constituents is taken to be $0.4\mu$m.}

\subsection{Composition}

We consider inhomogeneous aggregates. We take 30\% of the constituents of the aggregate to be composed of amorphous carbon and 70\% to be composed of silicate. {All constituents have the same size, so these, and any further abundances mentioned in the text, are volume fractions.} {When we assume that all of the Si-atoms are in the silicates, the carbon/silicate ratio we take corresponds to approximately half of the solar abundance of carbon atoms to be in the grains.} Of the 70\% silicate a fraction is crystalline while the rest is amorphous. Crystalline silicates display a very pronounced spectral structure, and part of our study is to see how this spectral structure is displayed for various fractions of crystalline silicates.

The polarizabilities are computed using Eq.~(\ref{eq:CDE}) with laboratory measurements for the refractive index as a function of wavelength. For the amorphous carbon we used measurements by \citet{1993A&A...279..577P}. Forsterite and enstatite are crystalline materials that have a refractive index different for the electric field oriented along the different crystallographic axis. We take this into account in the computations coupling the dipoles. The refractive index data was taken from \citet{Servoin} and \citet{1998A&A...339..904J} for the forsterite and the enstatite respectively.

For the polarizability of the amorphous silicate constituents we take a mixture of various amorphous silicates according to the composition derived for the interstellar amorphous silicates by \citet{2007A&A...462..667M}. This mixture is given in table~\ref{tab:Materials} and represents an average amorphous silicate composition of Mg$_{1.36}$Fe$_{0.13}$Na$_{0.05}$Al$_{0.05}$SiO$_{3.59}$.

The polarizabilities as functions of the wavelength for all different materials used in our computations are shown in Fig.~\ref{fig:alphas}.

\begin{figure*}[!t]
\centerline{\resizebox{13.85cm}{!}{\includegraphics{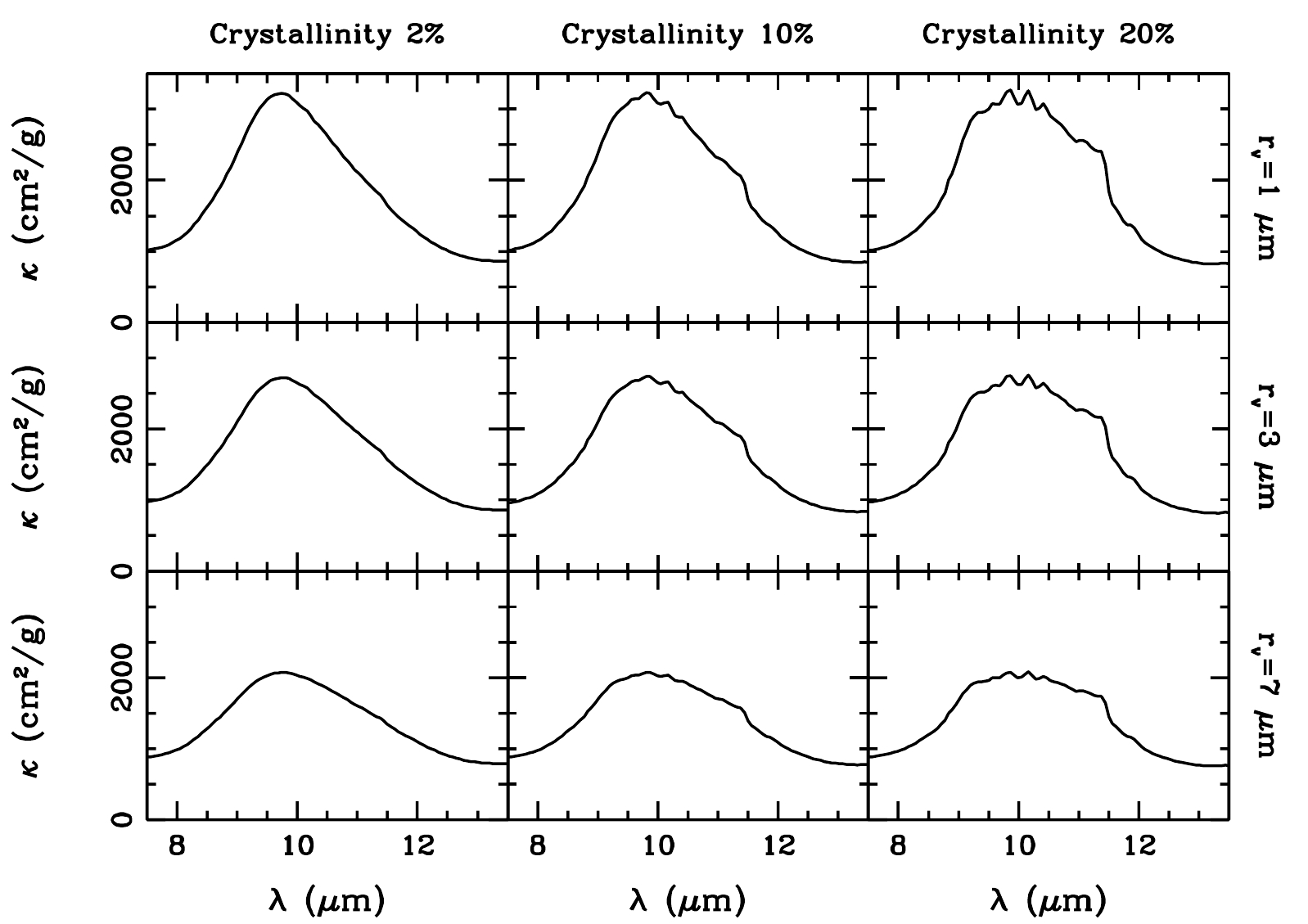}}}
\caption{The total mass aborption coefficients for the aggregates with crystalline silicate fraction (forsterite + enstatite) of 2\% (left panels), 10\% (middle panels), and 20\% (right panels). The volume equivalent radii of the aggregates are $1\,\mu$m (upper panels), $3\,\mu$m (middle panels), and $7\,\mu$m (lower panels).}
\label{fig:full spectra}
\end{figure*}

\section{Results}
\label{sec:results}

In Figs.~\ref{fig:full spectra} we plot the total mass absorption coefficient for the various aggregates. {Spectra are shown for various volume equivalent radii. By volume equivalent radius we mean the radius of a sphere with an equivalent volume as the aggregate.} We clearly see the expected flattening of the 10$\,\mu$m feature with increasing aggregate size. Also, for the more crystalline aggregates we see the spectral signature of the crystalline species appearing.

\begin{figure}[!t]
\resizebox{\hsize}{!}{\includegraphics{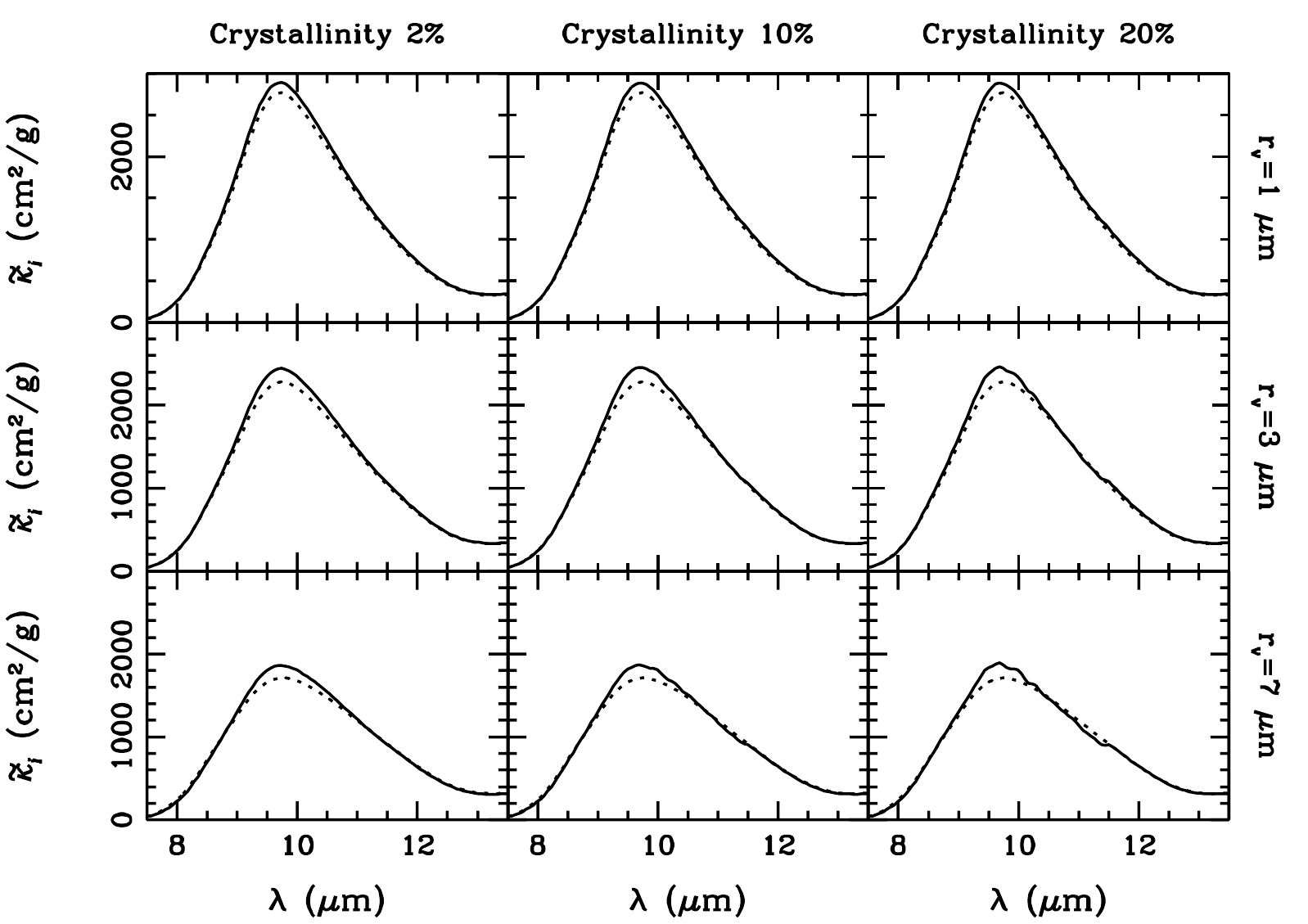}}
\caption{The effective mass absorption coefficient of the amorphous silicate component in the aggregates (full lines). The plots are organized as in Fig.~\ref{fig:full spectra}. As a reference we also plot the mass absorption coefficients for an aggregate with the same size but composed of pure amorphous silicate, {i.e. no carbon or crystalline silicates in the aggregate} (dashed line).}
\label{fig:ISM spectra}
\end{figure}

\begin{figure}[!t]
\resizebox{\hsize}{!}{\includegraphics{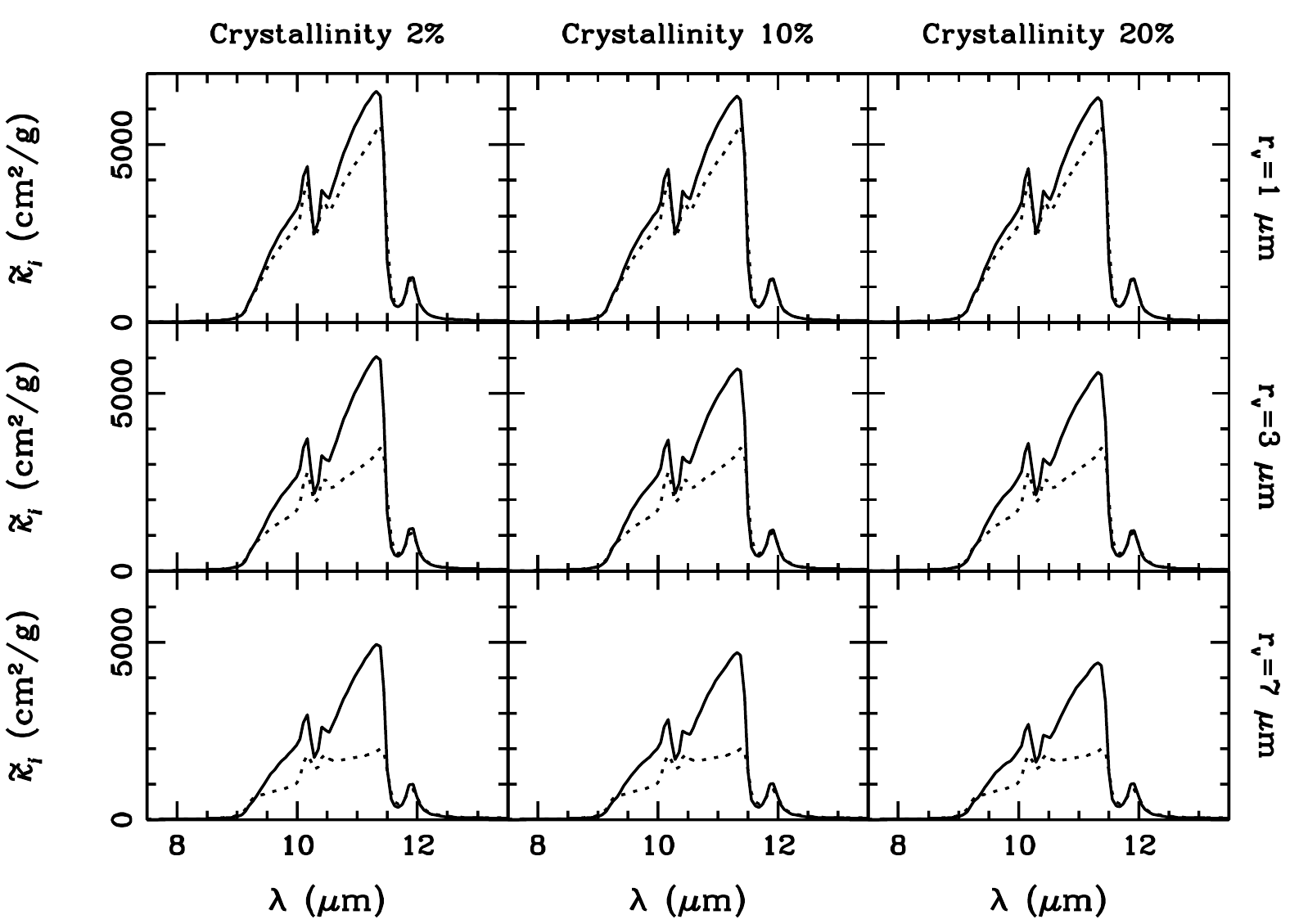}}
\caption{The effective mass absorption coefficient of the crystalline forsterite component in the aggregates (full lines). The plots are organized as in Fig.~\ref{fig:full spectra}. As a reference we also plot the mass absorption coefficients for an aggregate with the same size but composed of pure crystalline forsterite (dashed line).}
\label{fig:Forst spectra}
\end{figure}

In Figs.~\ref{fig:ISM spectra} and \ref{fig:Forst spectra} we plot the effective mass absorption coefficients of the amorphous silicate (Fig.~\ref{fig:ISM spectra}) and crystalline forsterite (Fig.~\ref{fig:Forst spectra}) components. In these plots we also show the mass absorption coefficients for the cases in which the entire aggregate is composed of a single material (amorphous silicate and crystalline forsterite respectively).

The abundance of amorphous silicate material in the aggregate is quite large. Therefore, it is not very surprising that for the amorphous silicate the pure and effective mass absorption spectra are quite similar (see Fig.~\ref{fig:ISM spectra}). For the forsterite component this is different. As can be seen from Fig.~\ref{fig:Forst spectra}, the effective mass absorption coefficients show a spectral signature typical for small grains even when the aggregate as a whole is large.

From these results we conclude that small inclusions of materials with strong resonances can show the spectral signature of small grains even when they are embedded in a larger grain. This is an important result to keep in mind when interpreting infrared spectra of environments where coagulated grains are expected to be present.

\begin{figure}[!t]
\resizebox{\hsize}{!}{\includegraphics{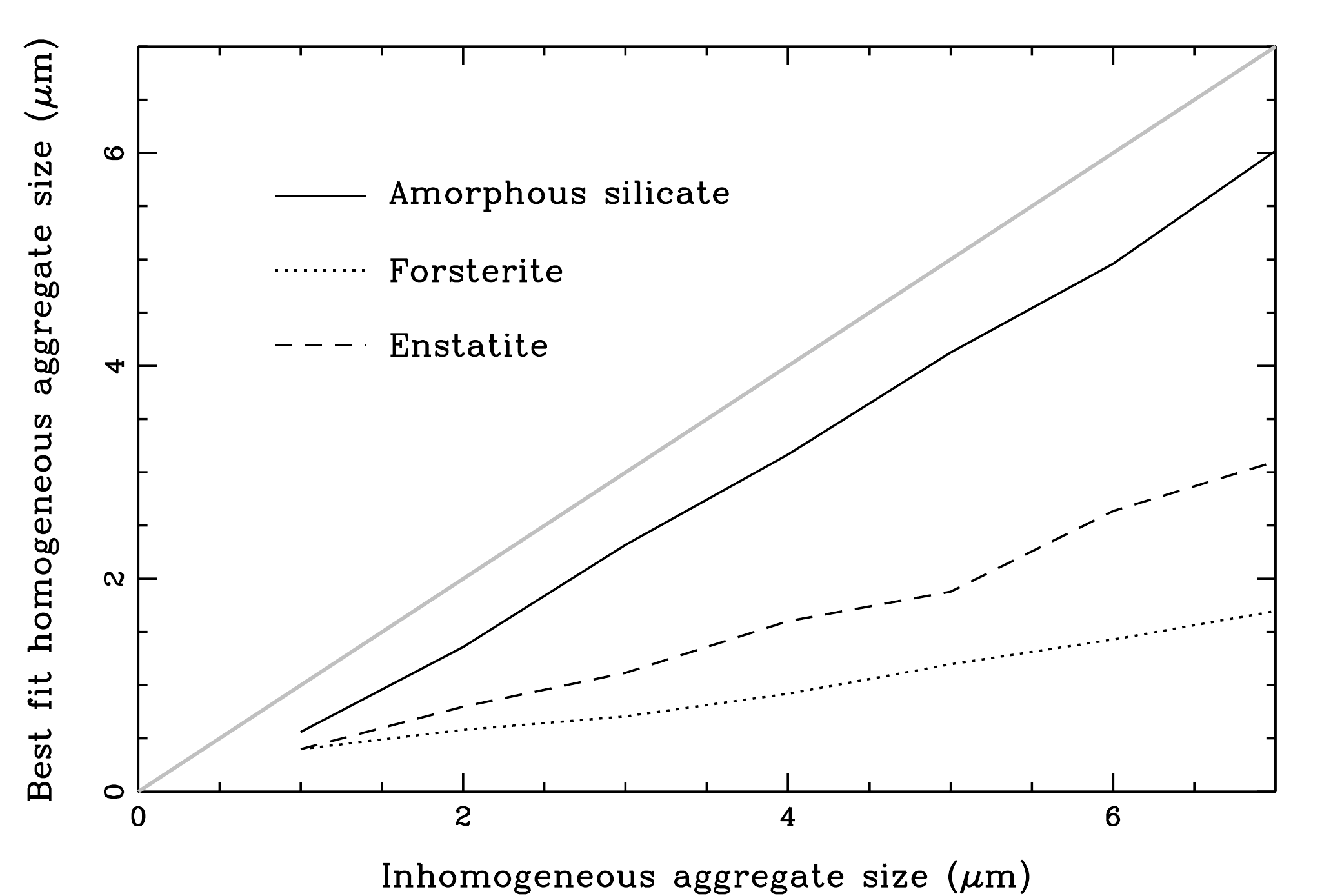}}
\caption{{The fitted apparent size for the different materials inside the aggregate as functions of the aggregate size for a crystallinity of 20\%. Results for crystallinities of 2 and 10\% show similar trends. Along the horizontal axis we plot the volume equivalent radius of the inhomogeneous aggregates. Along the vertical axis we plot the volume equivalent radius of the best fit homogeneous aggregate. For example, we can consider an inhomogeneous aggregate with the composition as given in the text, a crystallinity of 20\%, and a volume equivalent radius of $4\,\mu$m. From this figure we can now determine that this aggregate is best represented spectroscopically by a polydisperse mixture of \emph{homogeneous} aggregates with amorphous silicate aggregates having a volume equivalent radius of $r_V=3.3\,\mu$m, forsterite aggregates with $r_V=1.3\,\mu$m, and enstatite aggregates with $r_V=1.9\,\mu$m.}}
\label{fig:sizes fit}
\end{figure}

To quantify the statement above we determined the {apparent size} for the various components making up the aggregate. {This is the size that the components appear to have spectroscopically.} To do this we applied the following procedure.
\begin{enumerate}
\item {We computed the effective mass absorption coefficients, $\tilde{\kappa}_i$, of the various materials as function of wavelength.}
\item {We computed the mass absorption coefficients for \emph{homogeneous} aggregates, i.e. consisting of a single material, for all materials and for all sizes up to $7\,\mu$m for which we have computations (1, 2, 3, 4, 5, 6 and $7\,\mu$m).}
\item {The spectral shape of the effective mass absorption spectra computed in step 1 were compared with the mass absorption spectra for homogeneous aggregates computed in step 2 to determine the apparent sizes for all materials in the aggregate. In this comparison we also allowed a scaling factor for the spectra to be fitted. In this way we fit the spectral \emph{shape} and not to the absolute value of the mass absorption coefficient.}
\end{enumerate}
The results of this procedure are shown in Fig.~\ref{fig:sizes fit}. It is clear from this figure that the forsterite and enstatite grains appear much smaller than the amorphous silicate grains. This is caused by the small volume fraction of the crystalline materials.

Note that although the spectral \emph{shape} of the low abundance materials is representative for small grains, the strength is reduced compared to the mass absorption coefficient of individual small grains. This makes it very hard to get reliable abundance information when using single grains to analyze aggregate spectra. In the next section we will discuss an approximate method that can be used to correctly obtain abundance information from absorption or emission spectra caused by aggregate dust particles.

\section{Effective medium approach}
\label{sec:effective medium}

In this section we outline a fast, empirical, effective medium approach to simulate the absorption spectra of the aggregates discussed above. {In contrast to the CDA computations, which can take up to several days to obtain a spectrum, this method requires only Mie computations which can be done for any particle size in negligible computer time.} As we will show, this method almost perfectly reproduces the CDA computations while the computation time required is much smaller. The method we propose is based on the idea of simulating the entire aggregate as a sphere with an effective refractive index. There are several methods in the literature to compute the effective refractive index under certain assumptions. Recently \citet{2007ApOpt..46.4065V} have analyzed several of these effective medium theories for the optical properties of porous particles in the visual part of the spectrum. In our case the aggregates are very fluffy, open structures. If one would draw a circumscribing sphere around the aggregate the vacuum fraction inside this sphere would be very large. 

{The two most commonly applied mixing rules, the Garnett and Bruggeman mixing rules, can be derived from the same basic equation
\begin{equation}
\label{eq:mixing}
\alpha_a(m_\mathrm{eff}/m_m)=f_\mathrm{fill}\sum_{i=1}^{N} f_i\alpha_c(m_i/m_m)+(1-f_\mathrm{fill})\alpha_c(1/m_m),
\end{equation}
where $\alpha_a(m)$ and $\alpha_c(m)$ are the polarizabilities of the aggregate as a whole and the components of the aggregate respectively, $m_\mathrm{eff}$ is the effective refractive index of the material, $m_m$ is the so-called matrix material, $f_i$ is the volume fraction of component $i$, $m_i$ is its refractive index, and $f_\mathrm{fill}$ is the filling factor of the aggregate. Note that also vacuum is included as a component, with $m_i=1$ and volume fraction $(1-f_\mathrm{fill})$. The difference between the Garnett and the Bruggeman mixing rule is in the choice of the matrix material $m_m$. This refractive index represents the material in which all components are assumed to be suspended. In the Garnett rule the dominant component is used for $m_m$. However, when none of the components is clearly dominant, a better choice is to take $m_m=m_\mathrm{eff}$, so all components are assumed to be suspended in the \emph{effective medium itself}. This results in the Bruggeman mixing rule. Indeed, \citet{2007ApOpt..46.4065V} find that the Bruggeman mixing rule generally gives better results. However, when one of the materials clearly dominates, both mixing rules converge towards each other. In our case the filling factor of the aggregates is so low that vacuum is clearly the dominant component ($\sim$95\%). Therefore, we use the Garnett mixing rule taking $m_m=1$.
}

{In the classical form of the mixing rules both $\alpha_a$ and $\alpha_c$ are taken to be the polarizabilities of spheres. 
For irregularly shaped constituents we can adapt this by taking in Eq.~(\ref{eq:mixing}) the polarizability $\alpha_c$ to be that of an irregularly shaped particle. The aggregate as a whole is still approximately spherical so for $\alpha_a$ we take the polarizability of a sphere. Also substituting $m_m=1$ this results in
\begin{equation}
\label{eq:meffective}
3V\frac{m_\mathrm{eff}^2-1}{m_\mathrm{eff}^2+2}=f_\mathrm{fill}\sum_{i=1}^N f_i\,\alpha_c(m_i).
\end{equation}
In our case we use the CDE polarizability for the $\alpha_c$ as given by Eq.~(\ref{eq:CDE}). We note here that one could also have written the polarizability for a CDE grain on the left hand side, i.e. for $\alpha_a$. Since the aggregates are very fluffy structures $m_\mathrm{eff}$ is generally close to unity, so the effect of the circumscribing shape of the aggregates (which is what $\alpha_a$ represents) is generally very small. Thus,} in order to do computations for different aggregate sizes one might simply increase the size of the circumscribing sphere and perform Mie computations using the effective refractive index as given by Eq.~(\ref{eq:meffective}). We will refer to this method as the Aggregate Polarizability Mixing Rule (APMR).

\begin{figure*}[!t]
\centerline{\resizebox{13.85cm}{!}{\includegraphics{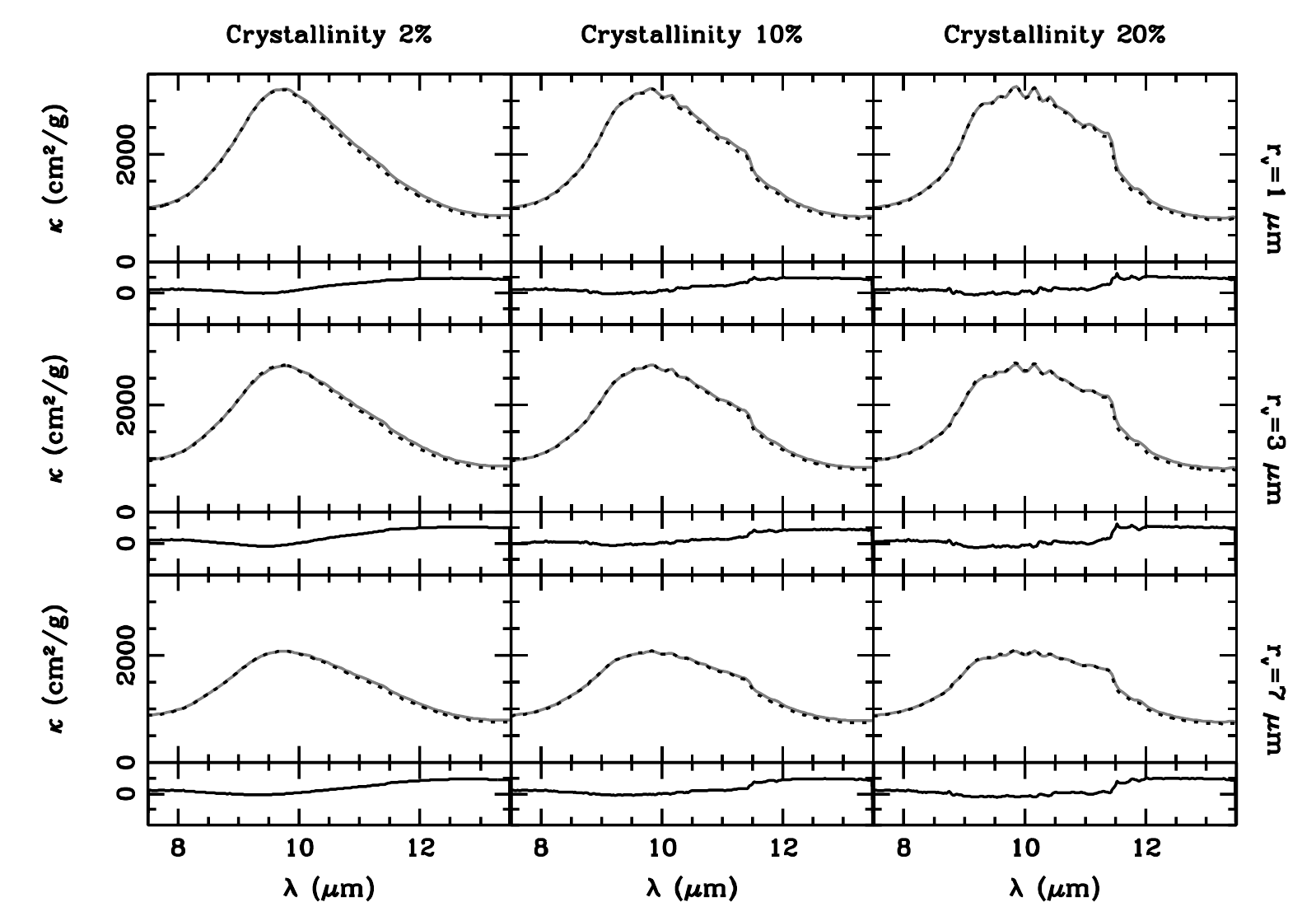}}}
\caption{A comparison of the spectra computed using the CDA computations and the effective medium approximation (APMR) for various values of the crystallinity (forsterite + enstatite volume fraction). The solid line represents the CDA computations while the dashed line respresents the resulting opacities from the effective medium approximation. Below each panel the relative error between the two curves is shown. Scale in these panels runs from -10\% to +10\%. The differences are in all cases smaller than 6\%.}
\label{fig:effective medium}
\end{figure*}

Now we have to determine how to get the required $f_\mathrm{fill}$ and radius of the homogeneous sphere. As is shown by \citet{2006A&A...445.1005M} the most natural choice, namely using the circumscribing sphere, leads to an overestimate of the result of the fluffyness of the aggregate. {Computations we performed for the case at hand confirm this conclusion.} This is because the aggregate constituents are not randomly distributed in the volume circumscribed by this sphere. Therefore, we have to choose a somewhat smaller radius which also somehow takes into account the structure of the aggregates. A choice, which leads to excellent results as shown below, is to use the radius of gyration of the aggregate.
\begin{equation}
\label{eq:rg}
r_g^2=\frac{1}{N}\sum_{i=1}^N \left|\vec{r}_i-\vec{r}_0\right|^2.
\end{equation}
In this equation $\vec{r}_i$ is the location of constituent $i$, and $\vec{r}_0$ is the center of mass of the aggregate. For fractal aggregates with fractal dimension $D_f$ the radius of gyration can be expressed as \citep{Filippov}
\begin{equation}
\label{eq:rg2}
r_g=\gamma r_V^{3/D_f},
\end{equation}
where $\gamma$ is a constant depending only on the size of the constituents and on the so-called fractal prefactor. By comparing the gyration radii computed using Eqs.~(\ref{eq:rg}) and (\ref{eq:rg2}) one finds that the aggregates we use are best represented using $D_f=2.82$ and $\gamma=2.44$.
The filling factor is now defined by
\begin{equation}
f_\mathrm{fill}=\frac{r_V^3}{r_g^3},
\end{equation}
where $r_V$ and $r_g$ are the volume equivalent radius and the gyration radius respectively. For the radius of the homogeneous sphere with the effective refractive index as given by Eq.~(\ref{eq:meffective}) in the final computation we now have to take $r_g$, which results in the aggregate and the sphere having the same mass.

{Substituting $m_m=m_\mathrm{eff}$, the Bruggeman rule for irregularly shaped constituents results in
\begin{equation}
\label{eq:Brug}
f_\mathrm{fill}\sum_{i=1}^N f_i\,\alpha_c(m_i/m_\mathrm{eff})+(1-f_\mathrm{fill})\alpha_c(1/m_\mathrm{eff})=0.
\end{equation}
The advantage of using the Garnett mixing rule over the Bruggeman rule is that it has an analytic solution for $m_\mathrm{eff}$ while the Bruggeman rule requires one to solve Eq.~(\ref{eq:Brug}) to find $m_\mathrm{eff}$. We have performed computations using the Bruggeman rule and find that in our case the differences with the Garnet rule according to Eq.~(\ref{eq:meffective}) are negligible. As explained above, this is because the aggregates we consider generally have a filling factor of less than 5\%. If one would like to extend the equations for the APMR to aggregates of different structures this is not necessarily the case. Considering the results of \citet{2007ApOpt..46.4065V} it is to be expected that especially for aggregates with a larger filling factor, i.e. lower porosity, it might be wise to use the Bruggeman mixing rule according to Eq.~(\ref{eq:Brug}) in stead of the Garnett rule (Eq.~\ref{eq:meffective}). However, in general aggregates formed in protoplanetary disks are expected to have fractal dimensions lower than the ones we used and thus be even more porous.}

With the simple equations above we can almost perfectly reproduce all our CDA computations. In Fig.~\ref{fig:effective medium} we compare the spectra for a mixture of 2, 10 and 20\% crystalline silicates computed using the CDA method outlined in section \ref{sec:computational approach} with the effective medium computations as outlined above. As one can see, the match is almost perfect.

\section{Conclusions}
\label{sec:conclusions}

We have studied the spectral behavior of inhomogeneous aggregates of irregularly shaped constituents. We find that the spectral appearance of such aggregates is a complex interplay between the abundances of the materials and the size of the aggregate. Also the shape and structure of the aggregate plays an important part in the spectral appearance of such particles.

The most striking conclusion of our computations is that the {apparent particle size of a dust component, i.e. the size derived from the infrared spectrum}, is a strong function of its abundance. Materials with a relatively low abundance will appear spectroscopically smaller than materials with a relatively high abundance.

We present a very efficient, fast and accurate approximate method to compute the opacities of the inhomogeneous aggregates using an empirical effective medium approach. We refer to this approach as the Aggregate Polarizability Mixing Rule (APMR). This approach can be used to study the effects of inhomogeneity of particles of larger sizes than can be computed using more exact methods given the large computational demand that these methods put on the computations for increasing particle size. 

Often, the analysis of infrared spectra of dusty environments is performed by assuming homogeneous, solid grains with various compositions and a limited number of grain sizes. 
The APMR approximate method is fast enough to be easily implemented in fitting algorithms trying to deduce the compositions and sizes of dust particles in these astrophysical environments. Furthermore, we propose to use the APMR to gauge the results obtained by previous studies and further investigate the effects of grain aggregation on the analysis of infrared spectra. Results of such an analysis will be presented in a separate paper.

\begin{acknowledgements}
We would like to thank A.~G. Hoekstra and M.~A. Yurkin for valuable discussions on the DDA. We also thank the referee, Nikolai Voshchinnikov, for useful comments. M.  Min acknowledges financial support from the Netherlands Organisation for Scientific Research (NWO) through a Veni grant.
\end{acknowledgements}

\end{document}